\documentstyle[aps,prb,twocolumn,graphicx]{revtex}
\begin{document}
\draft
\title{Oscillations of Andreev states in clean ferromagnetic films}
\author{M.~Zareyan$^{1,2}$, W.~Belzig$^1$, and Yu.~V.~Nazarov$^1$}
\address{
  $^1$ Department of Applied Physics and Delft Institute of
  Microelectronics and Submicrontechnology,\\ Delft University of
  Technology, Lorentzweg 1, 2628 CJ Delft, The Netherlands\\
  $^2$ Institute for Advanced Studies in Basic Sciences, 45195-159, 
  Zanjan, Iran}
\date{\today}
\maketitle
\begin{abstract}
  We investigate the influence of the exchange field on the Andreev bound
  states in a ferromagnetic (F) film backed on one side by a superconductor
  (S). Our model accounts for diffusive reflection at the outer surface and
  possible backscattering at the FS-interface. Phase shifting of the Andreev
  level by the exchange field results in an oscillatory behaviour of the
  density of states of F as a function of the layer thickness. We show that
  our results agree quantitatively with recent experiments.
\end{abstract}
\vspace*{5mm}
Probing the proximity effect by tunneling spectroscopy of induced
superconducting correlations has a long history. Early experiments on the
proximity density of states \cite{rowell} could be understood in the tunneling
model of McMillan\cite{mcmillan:68}. Recently the spatial dependence of the
proximity density of states in normal metals has been
measured\cite{pothier:96} and successfully explained in terms of the
quasiclassical theory\cite{belzig:96}.  The influence of spin-splitting by a
parallel magnetic field was measured in \onlinecite{tedrow} and found to
coincide with a Zeeman-split density of states.  A new experimental and
theoretical challenge is to extend these studies to the superconducting
proximity effect in ferromagnets.

Previous experimental investigation have concentrated on thermodynamic
properties of FS-multilayers. Here oscillations of the superconducting
critical temperature $T_{\text{c}}$ as a function of the thickness of the
F-layers have been predicted \cite{radovic} and studied experimentally
\cite{jiang}. However, the experimental evidence for these
$T_{\text{c}}$-oscillation is not conclusive. This may result from an
incomplete knowledge about the FS-interface quality.\cite{aarts1} In our
opinion, it is also questionable if the theoretical approach using the
diffusive quasiclassical formalism \cite{usadel} is applicable for F-layers of
typical thicknesses $~10-50$\AA.

The most recent experiments have concentrated on other properties of
FS-layers. Ryazanov {\em et al.} \cite{aarts} studied the supercurrent through
a thin ferromagnetic layer and found a non-monotonic temperature dependence,
which can be interpreted in terms of a $\pi$-phase shift due to the exchange
splitting. Kontos {\em et al.} \cite{aprili}, on the other hand, have studied
the density of states in a thin ferromagnetic layer in contact to a
superconductor. An oscillatory behaviour of the induced superconducting
correlation was observed for layers of different thicknesses and attributed to
influence of the exchange field. It is this experiment, that motivated our
present study.

Bearing this in mind, we study the superconducting proximity effect in a thin
ferromagnetic layer. The F-film is characterized by an homogenous exchange
splitting $h$. We model the film as a ballistic layer with rough boundaries.
Band mismatch and disorder at the interface may lead to enhanced
backscattering at the FS-boundary. We will derive a general formula for the
subgap density of states depending only on the length distribution of
classical trajectories in the F-layer. The resulting density of states shows
as a signature of the exchange splitting an oscillatory behaviour as a
function of layer thickness. Comparison to the experimental data show
reasonable agreement keeping in mind the large uncertainty of some sample
parameters.

\begin{figure}[htbp]
  \begin{center}
    \includegraphics[width=7.5cm]{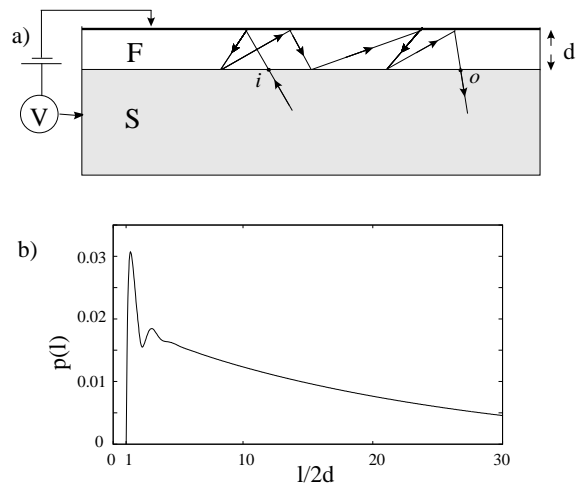}\\[2mm]
    \caption[]{
      a) Schematic drawing of a ferromagnetic film in connection with a
      superconductor.  A typical classical trajectory is also indicated. An
      electron coming from the bulk of the superconductor enters into the
      ferromagnet (at point $i$) and after several diffusive reflections from
      the insulator and the SF-interface returns to the superconductor (at
      point $o$). b) The calculated distribution of the trajectory lengths in
      the F-layer for small SF-transparency (here $T=0.1$). The double peak
      structure close to the smallest length originates from the first two
      reflections from the insulator, whereas the distribution for long
      trajectories decays as $\exp(-l/\bar{l})$, with the average length
      $\bar{l}\approx 2d/T$}
    \label{fig1}
  \end{center}
\end{figure}

The system we consider is sketched in Fig.~\ref{fig1}. A thin ferromagnetic
layer (F) of thickness $d$ is connected to a superconducting bank (S) on one
side and bounded on the other side by an insulator. The F-layer is
characterized by an exchange splitting, which we take into account as mean
field $h$ in the Hamiltonian.  The thickness $d$ is larger than the Fermi wave
length $\lambda_{{\text{F}}}$ and smaller than the elastic mean free path
$\ell_{\text{imp}}$ which allows for a quasi-classical
description\cite{eilenberger} in the clean limit. Then, the real-time
Eilenberger equation reads
\begin{eqnarray}
  \label{Eil}&&
  -i{\bbox{v}_{\text{F}}}
  \bbox{\nabla}\hat{g}_{\sigma}(E,{\bbox{v}_{\text{F}}},\bbox{r})=\\
  \nonumber&&
  \qquad\qquad\left[(E+\sigma h(\bbox{r}))\hat{\tau_3}
    -i\hat\tau_2\Delta(\bbox{r}) ,
  \hat{g}_{\sigma}(E,{\bbox{v}_{\text{F}}},\bbox{r})\right]\,.
\end{eqnarray}
Here $\hat{\tau}_i$ denote the Pauli matrices, $\Delta (\bbox{r})$ is the
(real) superconducting pair potential, and $\sigma$ ($=\pm 1$) labels the
electron spin. The matrix Green's functions have to obey the normalization
condition $\hat{g}^2_\sigma=1$. Inside the F-layer $h$ is constant and
$\Delta=0$.  We neglect the change of the pair potential in S, thus
$\Delta(\bbox{r}) =\text{const.}$ inside the superconductor.  We have solved
Eq.~(\ref{Eil}) along each classical trajectory in F that comes from the
superconductor and ends there. The density of states for $|E|<\Delta$ on a
given trajectory of length $l$ is then given by
\begin{eqnarray}
  \nonumber
  N(E,l)&=&\frac{N_0}{4}\sum\limits_{\sigma=\pm1}
  \text{Re}\left[\text{Tr}\hat\tau_3
    \hat{g}_\sigma(E,{\bbox{v}_{\text{F}}},\bbox{r})\right]\\
  \label{trajdos}
  &=&\frac{N_0}{2}\sum\limits_{\sigma=\pm1} 
  \frac{\pi v_{{\text{F}}}}{|E+\sigma h|} 
  \sum_{n=-\infty} ^{\infty}\delta (l-l_n),
\end{eqnarray}
where
\begin{equation}
  l_n=\frac{v_{{\text{F}}}}{E+\sigma h}\left(n\pi+\arccos(E/\Delta)\right). 
  \label{lnn}
\end{equation}
Here $N_0$ is the density of states at the Fermi level in the normal state.
Inside the F-layer $N_{\sigma}(E,l)$ is constant along a given trajectory and
depends only on the length of the trajectory $l$. Eq.~(\ref{trajdos}) means
that the density of states below $\Delta$ is a sum of $\delta$-peaks resulting
from Andreev bound states. The energies follow from the quasi-classical
quantization $l=l_n$.  The total density of states can then be found by
averaging (\ref{trajdos}) over all classical trajectories.  Denoting the
trajectory length distribution $p(l)$ we find for the density of states
\begin{eqnarray}
  \label{eq:dos}
  N(E)& =&\frac{N_0}{2}\sum\limits_{\sigma=\pm 1} 
  \frac{\pi v_{{\text{F}}}}{|E+\sigma h|} 
  \sum_{n=-\infty}^{\infty} p(l_n)\,.
\end{eqnarray}
This formula presents the general result for subgap density of states of a
quasiballistic ferromagnet connected to a superconductor. It is completely
specified by the length distribution of classical trajectories. This depends
only on the geometrical properties of the attached ferromagnet and the
connecting interface.

Now we have to specify the trajectory length distribution for our particular
case.  We model the thin layer by a weakly disordered thin film with a rough
surface and a rough SF-interface of average transparency $T$.  A typical
trajectory is depicted in Fig. 1. An electron coming from the bulk of S enters
into the F-layer and after several reflections from the insulator and the
FS-interface returns to the S-bank (see Fig.~\ref{fig1}), where it is Andreev
reflected as a hole which traverses the trajectory in the opposite direction.
Thus, the elementary building block of a typical trajectory is the segment
between two successive reflections from the superconductor. The number of
blocks which form the total trajectory depends on the transparency of the
interface, i.~e.~it is roughly $\sim 1/T$.  First, let us consider the length
distribution of one elementary block.  Due to the roughness of the insulator
and the FS-interface the quasi-particles undergo diffusive reflection from
these boundaries. Incoming and outgoing direction are completely uncorrelated.
Accounting for the weak bulk disorder we include a factor
$\exp(-l/\ell_{\text{imp}})$. The length distribution of one elementary block
is then given by
\begin{equation}
  p_0(l)= \frac{2d}{Cl^2}\left[\frac{l-2d}{l-d}+
  \frac{2d}{l}\ln\frac{l-d}{d}\right]e^{-l/\ell_{\text{imp}}}\theta(l/d-2),
  \label{p0}
\end{equation}
where $C=\text{E}_2^2(d/\ell_{\text{imp}})$ ($\text{E}_2(z) =
\int_{1}^{\infty}dx\exp{(-zx)}/x^2$ is the exponential integral of second
order).  Second, we connect the elementary building blocks, if the S-F
interface has an average transparency $T$. In determining the length
distribution we assume that an particle either goes through the interface or
is fully reflected. Only the number of these reflections depend on $T$. We do
not take into account quantum mechanical interference for a single reflection
at the FS-interface.  Taking this into account will yield essentially the same
results as our approach.  By an expansion in the reflectivity $R=1-T$ it is
easy to see that the full distribution $p(l)$ obeys the integral equation
\begin{equation}
  p(l)=Tp_0(l) +R\int dl^{\prime}p_0(l^{\prime})p(l-l^{\prime})\,.
  \label{pli}
\end{equation}
This is readily solved by a Fourier transformation. We obtain
\begin{equation}
  p(l)= \int_{-\infty}^{\infty} \frac{dk}{2\pi} e^{ikl} 
  \frac{TP_0(k)}{1-R P_0(k)},
  \label{plf}
\end{equation}
where $P_0(k)=\text{E}_2^2(ikd+d/\ell_{\text{imp}})/C$ is the Fourier
transform of $p_0(l)$.  The distribution $p(l)$ is plotted for $T=0.1$ and
$d/\ell_{\text{imp}}=0.1$ in Fig.~\ref{fig1}. It's characteristics are a
double peak structure close to the shortest trajectories $l\gtrsim 2d$
resulting from trajectories reflected once or twice from the insulator. At
large $l$ the distribution decays exponentially as $\exp(-l/\bar{l})$, where
${\bar l} =2d\ln(\ell_{\text{imp}}/d)/T$ is the mean trajectory length.  For
$T\sim 1$ $p(l)$ has only one peak close to $2d$. We therefore have two
characteristic lengths of the distribution, the smallest possible trajectory
length $2d$ and the average length $\bar l$. Which of these length scales will
determine the total density of states will depend on the other parameters, in
particular on $h$.

Combining Eqs. (\ref{trajdos}), (\ref{eq:dos}) and (\ref{plf}) we
obtain for the total DOS
\begin{equation}
  N(E)=\frac 12\sum_{\sigma=\pm} 
  \sum_{n=-\infty}^{\infty}\frac{TP_0(k_n)}{1-RP_0(k_n)} 
  e^{2ni\arccos E/\Delta },
  \label{dos}
\end{equation}
where $k_n=2n(E+\sigma h)/v_{\text{F}}$. Thus, the density of states is fully
expressed in terms of known functions. In the following we will discuss the
parameter range of rather strong exchange fields in the limit of thin layers
$d\ll v_{\text{F}}/\Delta$. The results presented below are in the most
realistic case of small SF-interface transparency $T\ll 1$. In this limit the
distribution of long lengths is well approximated as $\exp(-l/\bar l)$.  Most
probably the F-film has a non-uniform thickness due to the large scale
roughness of the boundaries. For a smoothly varying thickness we can take this
into account by averaging the (\ref{dos}) over a Gaussian distribution of the
thicknesses around a mean value $d$. This will also lead to a smoothening of
the sharp features in the DOS resulting from the lower cutoff in $p(l)$. The
qualitative behavior will however not change. In practice, we have chosen a
width of the distribution to be of order $10\%$ that corresponds to condition
of the experiment \cite{aprili}.

\begin{figure}[htbp]
  \begin{center}
   \vspace*{0.7cm}
    \includegraphics[width=7.cm]{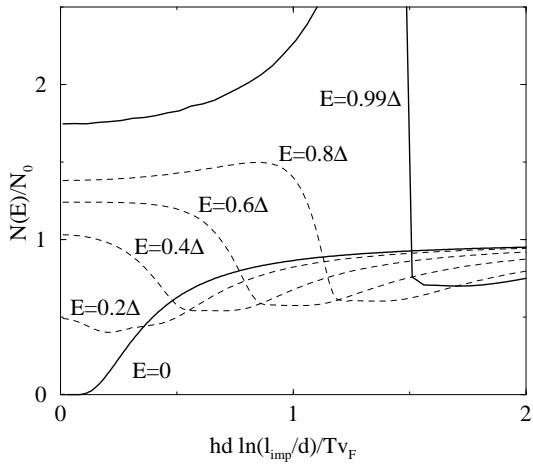}
    \vspace*{0.5cm}
    \caption{
      Density of states for weak exchange fields and small SF-interface
      transparency. For $E=0$ the curve is independent of $T$ as long as $T\ll
      1$. At finite energies the precise form depends on the transparency
      (here $T=0.03$ and $\ell_{\text{imp}}=100 d$). It has a broad peak at
      zero energy and two dips $E=\pm h$ resulting from the shifts of the
      Andreev bound states.}
 \label{fig2}
  \end{center}
\end{figure}

For weak exchange fields when $h \bar l/v_{\text{F}} \lesssim 1$ the DOS at
low energies $E\ll \Delta$ is mainly contributed from long trajectories with
$l \gtrsim \bar{l}$.  The DOS versus $ h d /T$ is plotted for different
energies in Fig 2.  For $h=0$ (a normal film), $N(E)$ vanishes at zero energy
and increases for finite energies ending with a peak at $\Delta$. The effect
of the exchange field in the regime $h<\Delta$ is to split the DOS for the two
spin bands. Thus the total DOS is the average of two by $\pm h$ shifted spectra
resulting in a peak at zero energy and two dips at $E=\pm h$. Increasing $h$
further leads to a suppression of superconducting features of the DOS.  The
zero energy DOS increases roughly as $(\pi v_{\text{F}}/h{\bar l}) \exp(-\pi
v_{\text{F}}/2h{\bar l})/(1-\exp(-\pi v_{\text{F}}/h{\bar l}))$, which follows
from the approximation mentioned above. At finite energies $N(E)$ passes
through minima corresponding to $E=\pm h$ before approaching the normal metal
DOS.

For larger exchange fields when $hd\gtrsim 1$ mainly the short trajectories of
$l\sim 2d$ contribute to the energy dependence of $N(E)$. The DOS is close to
that of the normal state.  The interesting part are the small deviations of
amplitude $T$ which oscillate as a function of $dh$. In Fig. 3 we have plotted
$(N(E)/N_0-1)/T$ versus $dh/v_{\text{F}}$ for different energies.  At small
$hd/v_{\text{F}} \ll 1$ and $E\ll\Delta$ the DOS as a function of
$hd/v_{\text{F}}$ still reminds of the double peak structure of the length
distribution. For $hd/v_{\text{F}} \gtrsim 1$, however, it develops coherent
oscillations as a function of $hd/v_{\text{F}}$ with a period of $\pi/2$. Note
that the magnitude and the sign of the oscillation depend on the energy.
Maximal amplitudes of opposite sign always correspond to $E=0$ or
$|E|=\Delta$.

\begin{figure}[htbp]
  \begin{center}
    \includegraphics[width=7cm]{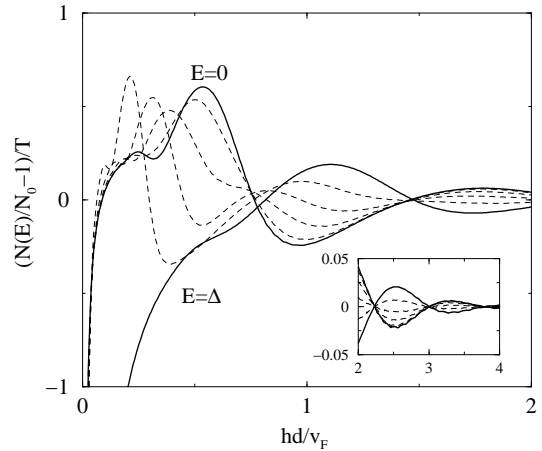}
    \caption[]{ 
      Oscillation of reduced DOS $(N(E)/N_0-1)/T$ with the exchange field. For
      $hd/v_{\text{F}} \gtrsim 1$, it develops coherent oscillations as a
      function of $dh/v_{\text{F}}$ with a period of $\pi/2$ and an energy
      dependent magnitude and sign. Maximal amplitudes of opposite sign always
      correspond to $E=0$ or $|E|=\Delta $.}
    \label{fig3}
  \end{center}
\end{figure}

Let us now discuss the relation of our results to the recent experiment
performed in the group of Aprili \cite{aprili}. They observed DOS oscillation
in thin films of the ferromagnetic alloy Pd$_{1-x}$Ni$_{x}$ with $x$ of the
order of 10\% in contact with a Nb electrode.  By tunneling spectroscopy they
measured the differential conductance versus bias of the F-film at low
temperatures $\sim 300$mK. Normalized to the normal state conductance this
yields the density of states in the ferromagnetic layer. For two different
thicknesses of the F-film, $d_1=50${\AA} and $d_2=75${\AA} the reduced density
of states differs in sign and magnitude, but has the same functional energy
dependence.  The exchange field was estimated by several methods to be in the
range $h=5-15$meV. These experimental determinations were, however, only
sensitive to the average magnetization. In thin layers, it is reasonable to
assume that the magnetic structure has multiple domains, inhomogeneous
thickness (as already discussed before), and non-uniform Ni concentration. All
these effects tend to suppress the measured values of the exchange field below
the value, which results from the local Ni-doping level. We will therefore
regard the experimental values of $h$ as lower bounds to the value used in our
fits.  In Fig. 4 we compare the data of Ref.~\cite{aprili} with our
calculation.  We have taken $v_{\text {F}}=2\times 10^{7}{\text {cm/s}}$ and
$h=29$meV, and $T=0.055$.  The agreement between the experimental data and our
calculation is satisfactory. Both, the functional dependence and the sign
change are correctly reproduced by our calculations.

\begin{figure}[htbp]
  \begin{center}
   \includegraphics[width=7cm]{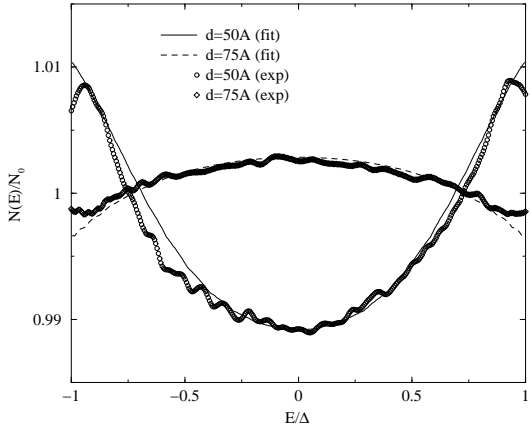}
    \caption[]{
      Energy dependence of DOS for two different thicknesses of the
      ferromagnetic film, $d_1=50$\AA\ (solid line) and $d_2=75$\AA\ (dashed
      line), with the corresponding experimental curves of Ref.~\cite{aprili}.
      The DOS oscillation appears as an inversion of the energy dependence of
      DOS when $d$ changes by $\pi v_{\text {F}}/4h$. }
    \label{fig4}
  \end{center}
\end{figure}

In conclusion we have investigated theoretically the superconducting proximity
effect in thin ferromagnetic layers in a quasi-ballistic model. We have found
that the effect of the ferromagnet exchange field $h$ is to suppress the
superconducting features in the density of states for $dh/v_{\text{F}}\gtrsim
T$.  At exchange fields larger than $qv_{\text{F}}/d$ the density of states
oscillates around the normal state value as a function of $dh/v_{\text{F}}$
with a period of $\pi/2$ and an amplitude of the order of the interface
transparency $T$. We have shown that the oscillation can lead to an inverted
energy dependence of the density of states. This effect has been observed in
the experiment of Ref.~\cite{aprili}. We have achieved quantitative agreement
between our theory and the experimental data.

We thank M.~Aprili and T.~Kontos for sharing their data with us prior to
publication, which was the initiation of this work. W.B. was financially
supported by the ``Stichting voor Fundamenteel Onderzoek der Materie'' (FOM)
and a Feodor Lynen Fellowship of the ``Alexander von Humboldt-Stiftung''. M.Z.
gratefully acknowledges financial support from the Iranian Ministry of
Science, Research and Technology and the TU Delft.

\end{document}